\documentclass[conference]{IEEEtran}
\ifCLASSINFOpdf
\usepackage[pdftex]{graphicx}
\else
\fi
\usepackage[cmex10]{amsmath}
\usepackage{algorithmic}
\usepackage{array}
\usepackage{mdwmath}
\usepackage{mdwtab}
\usepackage{eqparbox}
\usepackage[tight,footnotesize]{subfigure}
\usepackage{fixltx2e}
\usepackage{stfloats}
\usepackage{url}
\usepackage{graphicx}
\usepackage{mathrsfs}
\usepackage{caption2}
\begin{document}
\title{Asynchronous Physical-layer Network Coding Scheme for Two-way OFDM Relay}


\author{\IEEEauthorblockN{Xiaochen Xia, Kui Xu, Youyun Xu}
\IEEEauthorblockA{Institute of Communication Engineering, PLAUST\\
Nanjing 210007, P. R. China\\
Email: Xia1382084@gmail.com}}


%Telephone: (800) 555--1212, Fax: (888) 555--1212}


\maketitle

\begin{abstract}
%\boldmath
In two-way OFDM relay, carrier frequency offsets (CFOs) between relay and terminal nodes introduce severe inter-carrier interference (ICI) which degrades the performance of traditional physical-layer network coding (PLNC). Moreover, traditional algorithm to compute the \emph{posteriori} probability in the presence of ICI would incur prohibitive computational complexity at the relay node. In this paper, we proposed a two-step asynchronous PLNC scheme at the relay to mitigate the effect of CFOs. In the first step, we intend to reconstruct the ICI component, in which space-alternating generalized expectation-maximization (SAGE) algorithm is used to jointly estimate the needed parameters. In the second step, a channel-decoding and network-coding scheme is proposed to transform the received signal into the XOR of two terminals' transmitted information using the reconstructed ICI. It is shown that the proposed scheme greatly mitigates the impact of CFOs with a relatively lower computational complexity in two-way OFDM relay.

\end{abstract}


% \fi
\IEEEpeerreviewmaketitle

\section{Introduction}
Nowadays, there is increasing interest in employing the idea of network coding [1] in wireless communication to improve the system throughput [2]-[5]. The simplest scenario in which network coding can be applied is the two-way relay channel (TWRC), as illustrated in Fig. 1. In TWRC, two terminal nodes $T_1$ and $T_2$ exchange statistically independent information with the help of a relay node $R$. Traditionally, this process can be achieved within four time slots, that is, $T_1\to R$, $R\to T_2$, $T_2\to R$ and $R\to T_1$, as illustrated in Fig. 1(a). To enhance the system throughput of TWRC, physical-layer network coding (PLNC) has been introduced in [6]. PLNC reduces the required time slots for one round of information exchange from four to two comparing with the traditional protocol, as shown in Fig. 1(b).

In this paper, we consider the OFDM modulated TWRC or two-way OFDM relay (TWOR). A key issue in practical application of PLNC in TWOR is how to deal with the frequency asynchrony between the signals transmitted by the two terminal nodes. That is, symbols transmitted by different terminals may arrive at the relay node with different CFOs. Due to the impact of CFOs, traditional channel-decoding and network-coding mapping method in [6] suffers from severely performance degradation. Moreover, traditional algorithm [7] to compute the \emph{posteriori} probability at the relay node may introduce prohibitively expensive computation for practical implementation due to correlations among the received samples caused by ICI. On the other hand, the OFDM modulated PLNC assigns the same subcarrier to both terminals which is very different from OFDMA where the subcarriers of different users are orthogonal. That is to say, the received signal in each subcarrier is the composition of symbols transmitted by $T_1$ and $T_2$. Due to this observation, traditional CFO compensation methods developed for OFDMA [8][9] are difficult to be utilized in the PLNC system. In [10], Lu investigates the frequency asynchronous PLNC for OFDM system and proposes a method to compensate the CFOs with the mean of two terminals' estimated CFOs at the relay node. Unfortunately, this scheme will not perform well when the relative CFO between the two terminals becomes larger.

In this paper, we develop a two-step asynchronous PLNC scheme at the relay node. Comparing with the previous work: 1) The proposed method can effectively mitigate the effect of frequency offsets in TWOR system; 2) It can cope with the situation that the relative CFO is larger without incurring severe performance degradation with respect to perfectly synchronized system; 3) The proposed scheme has a relatively lower computational complexity.
\begin{figure}[t]
\centering
\includegraphics[width=8.5cm]{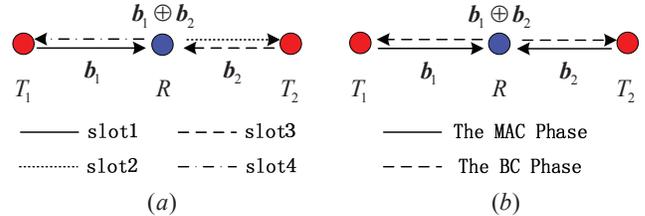}
\caption{A system model for two-way relay channel: (a) Traditional four-slot protocol; (b) PLNC.} \label{fig:graph}
\end{figure}

\emph{Notation}: Lower and upper case bold symbols denote column vectors and matrices, respectively. $(\cdot)^T$, $(\cdot)^*$, and $(\cdot)^H$ denote transpose, complex conjugate, and Hermitian transpose, respectively. $E\left(\cdot\right)$ stands expectation operation. $\textrm {diag}\left(\cdot\right)$ denote a diagonal matrix. Let $\hat{\vartheta}$ denote the estimate of ${\vartheta}$. Let $|\cdot|$ and $\|\cdot\|$ denote the magnitude and the Euclidean norm, respectively. We use $\boldsymbol I_N$ and $\boldsymbol 0_{N\times M}$ for the $N\times N$ identity matrix and $N\times M$ matrix with all zero entries, respectively.

\section{System Model}
We consider the TWOR network as shown in Fig. 1, where $T_1$ and $T_2$ exchange statistically independent information with the help of node $R$. It is assumed that all nodes are half-duplex, that is, a node cannot transmit and receive simultaneously. It is also assumed that each node is equipped with a single antenna and no direct link is existed between $T_1$ and $T_2$.

We consider the two-phase transmission scheme which consists of a multiple-access (MAC) phase and a broadcasting (BC) phase as illustrated in Fig. 1(b). During the MAC phase, terminals $T1$ and $T2$ send OFDM modulated signals to the relay node $R$ simultaneously. Let $\boldsymbol b_i$, $\boldsymbol c_i$ and $\boldsymbol u_i$ denote the uncoded source vector, channel coded vector and the modulated vector of terminal $T_i$, respectively. Let $N$ denote the total number of subcarriers. Let ${\boldsymbol x}_i(n)$ denote the $n$th frequency domain OFDM block of node $T_i$, where ${\boldsymbol  x}_i(n)=[x_{i,0}(n),x_{i,1}(n),\cdots,x_{i,N-1}(n)]^T$, $i\in\{1,2\}$. We define $\boldsymbol A_i$ as the subcarrier allocation matrix,\\
\begin{equation}
\left[\boldsymbol A_i\right]_{q,k}=\left\{\begin{array}{ll}
1 & \textrm{if the $q$th subcarrier is allocated to the $k$th}\\
 &\textrm{element of $\boldsymbol u_i(n)$}\\
0 & \textrm{if the $q$th subcarrier is not allocated to $T_i$}\end{array}\right..
\end{equation}
Then we have ${\boldsymbol x_i}\left( n \right) = {\boldsymbol A_i}{\boldsymbol u_i}\left( n \right)$. Notably, during the MAC phase, $T_1$ and $T_2$ are allocated a same subset of $K$ subcarriers due to the application of PLNC, so we can obtain that $\boldsymbol A_1=\boldsymbol A_2=\boldsymbol A$. Let ${\boldsymbol h}_i=[h_i(0),h_i(1),\cdots,h_i(L_i-1),{\boldsymbol 0}_{\left(N-L_i\right) \times 1}^T]^T$ denote the channel impulsive response (CIR) between $T_i$ and the relay node. Here we assume that the length of cyclic prefix (CP) $N_g\ge\max\{L_1,L_2\}$ to avoid the inter-block interference (IBI). Therefore, we concentrate only on the $n$th OFDM block and omit the index $n$ in the rest of this work. Then the received signal samples at node $R$ in the end of the MAC phase can be expressed as\\
\begin{equation}
{\boldsymbol y}_R={\boldsymbol E}(\varepsilon_1){\boldsymbol F}{\boldsymbol X}_1{\boldsymbol D}{\boldsymbol h}_1+{\boldsymbol E}(\varepsilon_2){\boldsymbol F}{\boldsymbol X}_2{\boldsymbol D}{\boldsymbol h}_2+\boldsymbol w,
\end{equation}
in which\\
$\bullet$ $\boldsymbol E(\varepsilon_i)=\textrm {diag}\{1,e^{j2\pi\varepsilon_i/N},\cdots,e^{j2\pi(N-1)\varepsilon_i/N}\}$ and $\varepsilon_i$ is the normalized CFO for node $T_i$;\\
$\bullet$ $\boldsymbol F$ is an $N\times N$ matrix with elements $[\boldsymbol F]_{p, q}=\sqrt{1/N}e^{j2{\pi}pq/N}$ for $0\le p,q\le N-1$;\\
$\bullet$ ${\boldsymbol X}_i=\textrm {diag}\{x_{i,0},x_{i,1},\cdots,x_{i,N-1}\}^T$ is a diagonal matrix;\\
$\bullet$ $\boldsymbol D$ is a Fourier matrix with elements $[\boldsymbol D]_{p, q}=e^{-j2{\pi}pq/N}$ for $0\le p,q\le N-1$;\\
$\bullet$ ${\boldsymbol w}=[w(0),w(1),\cdots,w(N-1)]^T$ is the additive white Gaussian noise vector with zero mean and covariance matrix $\sigma^2_w\boldsymbol I_N.$

By multiplying both sides of (2) by matrix $\boldsymbol F^H$, we obtain the frequency domain received samples which can be expressed as
\begin{equation}
\begin{aligned}
\boldsymbol Y_R&=\boldsymbol A^T\left(\boldsymbol {\Pi}_1\boldsymbol S_1+\boldsymbol {\Pi}_2\boldsymbol S_2\right)+\boldsymbol W\\
&{}=\underbrace{\boldsymbol A^T\left(\boldsymbol {\Lambda}_1\boldsymbol S_1+\boldsymbol {\Lambda}_2\boldsymbol S_2\right)}_{\textrm{Desired signal}}\\
&{}+\underbrace{\boldsymbol A^T\left(\left(\boldsymbol {\Pi}_1-\boldsymbol {\Lambda}_1\right)\boldsymbol S_1+\left(\boldsymbol {\Pi}_2-\boldsymbol {\Lambda}_2\right)\boldsymbol S_2\right)}_{\textrm{ICI}}+\boldsymbol W,
\end{aligned}
\end{equation}
in which $\boldsymbol {\Pi}_i=\boldsymbol F^H\boldsymbol E(\varepsilon_i)\boldsymbol F$ is defined as the interference matrix, $\boldsymbol {\Lambda}_i=\textrm {diag}\left[\boldsymbol {\Pi}_i\right]$ and $\boldsymbol S_i={\boldsymbol X}_i{\boldsymbol D}{\boldsymbol h}_i$. Obviously, $\boldsymbol {\Lambda}_i=\boldsymbol I_N$ and $\left(\boldsymbol {\Pi}_i-\boldsymbol {\Lambda}_i\right)=\boldsymbol 0_{N\times N}$ for synchronous case. In (3), we can see that, with non-zero CFOs, each output symbol is affected by ICI from all other subcarriers due to the loss of orthogonality among subcarriers. This results in poor performance for traditional channel-decoding and network coding mapping method [6].

Using the proposed mapping scheme for asynchronous PLNC detailed in the next section, relay node $R$ transforms the received superimposed signal in the presence of ICI into the XORed massages ${\boldsymbol b}_1\oplus{\boldsymbol b}_2={\rm T}\left(\boldsymbol Y_R\right)$. After that, in the BC phase, relay then broadcasts ${\boldsymbol b}_1\oplus{\boldsymbol b}_2$. Both $T_1$ and $T_2$ try to decode ${\boldsymbol b}_1\oplus{\boldsymbol b}_2$ from their corresponding received signals. Since $T_1\left(T_2\right)$ knows its own bits, after decoding ${\boldsymbol b}_1\oplus{\boldsymbol b}_2$, it can extract the bits transmitted by $T_2\left(T_1\right)$ from the XORed massages by subtracting its own information.

\section{Proposed Scheme}
%As shown in (3), the biggest difference between the OFDM modulated PLNC and OFDMA system is that the PLNC assigns the same subcarriers to both terminals. That is to say, the received signal in each subcarrier is the composition of symbols transmitted by $T_1$ and $T_2$ for PLNC system. Due to this observation, traditional CFO compensation methods developed for OFDMA [12][13] are difficult to be utilized in the PLNC system even when CFOs are perfectly known in the relay.
%To deal with this problem,
For frequency asynchronous PLNC, a critical challenge is how to map the received signal at the relay node into the XOR of two terminals' transmitted information. In this section, we present a two-step asynchronous PLNC scheme to deal with this problem. In the first step, we intend to reconstruct the ICI component in (3), in which SAGE algorithm is employed to jointly {estimate\footnotemark[1]}  $\boldsymbol \varepsilon=\left[\varepsilon_1,\varepsilon_2\right]^T$, $\boldsymbol h=\big[\boldsymbol h_1^T,\boldsymbol h_2^T\big]^T$ and $\boldsymbol X=\left[\boldsymbol X_1^T, \boldsymbol X_2^T\right]$. Here we suppose a coarse CFO compensation has been operated before the uplink frame, as a result, we need to concentrate only on the situation that CFOs are less than half of the subcarrier spacing, i.e., $-1/2<\varepsilon_i<1/2$, for $i=1,2$. Secondly, using the reconstructed ICI, an channel-decoding and network-coding scheme for asynchronous PLNC is performed to map the received signal into the XOR of two terminals' transmitted information.

\subsection{SAGE Based ICI Reconstruction}
Let $\boldsymbol \theta=\left[\boldsymbol \varepsilon^T, \boldsymbol h^T, \boldsymbol X^T\right]^T$ denote a set of parameters to be estimated from the observed data $\boldsymbol y_R$ with conditional probability density function $p(\boldsymbol y_R|\boldsymbol \theta)$. Obviously, the maximization problem of $p(\boldsymbol y_R|\boldsymbol \theta)$ with respect to the unknown parameters $\boldsymbol \theta$ is equivalent to the maximization of the log-likelihood function which is given by\\
\begin{equation}
\begin{aligned}
L(\boldsymbol \theta)=&- \frac{1}{{\sigma _w^2}}{\left\| \boldsymbol{y_R} -\boldsymbol E\left( {{{\boldsymbol\varepsilon }_1}} \right)\boldsymbol F{{ {\boldsymbol X}}_1}\boldsymbol D{{ {\boldsymbol h}}_1} - \boldsymbol E\left( {{{ {\boldsymbol\varepsilon} }_2}} \right)\boldsymbol F{{ {\boldsymbol X}}_2}\boldsymbol D{{ {\boldsymbol h}}_2} \right\|^2}\\
&{}+ const.
\end{aligned}
\end{equation}
Then we should consider parameter estimation from the viewpoint of maximizing $L(\boldsymbol \theta)$, that is\\
\begin{equation}
\hat{\boldsymbol \theta}=\arg \mathop {\max }\limits_{\boldsymbol \theta}L(\boldsymbol \theta).
\end{equation}

{\footnotetext [1]{The reason why we update the CFOs during the payload is two-fold: i) It is necessary to estimate the residual CFO due to the estimation error in the preamble; ii) For scenarios with time-varying CFOs, reconstructing the ICI with estimates from the preamble may results in poor performance.}}

However, direct computation of the maximization problem would require an exhaustive search over multiple-dimensional space spanned by $\boldsymbol \varepsilon=\left[\varepsilon_1,\varepsilon_2\right]^T$, $\boldsymbol h=\big[\boldsymbol h_1^T,\boldsymbol h_2^T\big]^T$ and $\boldsymbol X=\left[\boldsymbol X_1^T, \boldsymbol X_2^T\right]$, which may incur prohibitively expensive computation for practical implementation. To reduce the computational complexity, we propose a SAGE based scheme to estimate the multiple-dimensional parameters iteratively.

To operate the SAGE algorithm for asynchronous PLNC, we should divide the parameters to be estimated into two groups of $\boldsymbol \theta_i=\left[\varepsilon_i,\boldsymbol h_i^T,\boldsymbol X_i^T\right]^T$, for $i=1,2$. A hidden space [11] must be chosen for each group so that the update process of one group can be taken place while the other is kept fixed at its latest value. Here we define the hidden space as\\
\begin{equation}
{\boldsymbol y}_i={\boldsymbol E}\left(\varepsilon_i\right){\boldsymbol F}{\boldsymbol X}_i{\boldsymbol D}{\boldsymbol h}_i+\boldsymbol w.
\end{equation}
In (6), we include all the noise to the hidden space of $\boldsymbol \theta_i$. [11] has shown that such a choice is optimal to reduce the Fisher information and increase the convergence rate.

The update process of $\boldsymbol \theta_i$, $\forall i\in\left\{ {1,2} \right\}$, at $(m+1)$th iteration can be described as follow:

\textit {1) Expectation--step}: In this step, we define the conditional log-likelihood function [11] or $Q$ function of $\boldsymbol \theta_i$, that is\\
\begin{equation}
Q\left(\boldsymbol \theta_i\big|\hat{\boldsymbol \theta}^{[m]}\right)\buildrel \Delta \over = E\left\{\log p\left(\boldsymbol y_i\big|\boldsymbol \theta_i,\hat{\boldsymbol \theta}_{\bar i}^{[m]}\right)|\boldsymbol y_R,\hat{\boldsymbol \theta}^{[m]}\right\},
\end{equation}
in which $p\left(\boldsymbol y_i\big|\boldsymbol \theta_i,\hat{\boldsymbol \theta}_{\bar i}^{[m]}\right)$ is the conditional probability density function of $\boldsymbol y_i$,\\
\begin{equation}
\begin{aligned}
&p\left(\boldsymbol y_i\big|\boldsymbol \theta_i,\hat{\boldsymbol \theta}_{\bar i}^{[m]}\right)=p\left(\boldsymbol y_i\big|\boldsymbol \theta_i\right)\\
&{}=\frac{1}{(\pi\sigma_w)^N}\exp\left\{-\frac{1}{\sigma_w^2}\left\|\boldsymbol y_i-\boldsymbol E\left(\varepsilon_i\right)\boldsymbol F\boldsymbol X_i\boldsymbol D\boldsymbol h_i\right\|^2\right\}.
\end{aligned}
\end{equation}
where $\bar i\buildrel \Delta \over = \left\{1,2\right\}/\left\{i\right\}$. Substitute (8) into (7) and remove the terms that do not relate to $\boldsymbol \theta_i$, we can rewrite (7) as\\
\begin{equation}
Q\left(\boldsymbol \theta_i\big|\hat{\boldsymbol \theta}^{[m]}\right)=\textrm{Re}\left\{\left(\hat{\boldsymbol y}_i^{[m]}\right)^H\boldsymbol E\left(\varepsilon_i\right)\boldsymbol F\boldsymbol X_i\boldsymbol D\boldsymbol h_i\right\},
\end{equation}
in which $\hat{\boldsymbol y}_i^{[m]}$ is the estimate of $\boldsymbol y_i$ at the $(m+1)$th iteration and\\
\begin{equation}
\hat{\boldsymbol y}_i^{[m]}=\boldsymbol y_R-\boldsymbol E\left(\hat\varepsilon_{\bar i}^{[m]}\right)\boldsymbol F\hat{\boldsymbol X}_{\bar i}^{[m]}\boldsymbol D\hat{\boldsymbol h}_{\bar i}^{[m]}.
\end{equation}

\textit {2) Maximization--step}: In this step, we update the value of $\varepsilon_i$, $\boldsymbol h_i$ and $\boldsymbol X_i$ sequentially. The channel estimation at the $(m+1)$th iteration can be obtained by maximizing (9) with respect to $\boldsymbol h_i$ while fixing $\hat{\varepsilon}_i$ and $\hat{\boldsymbol X}_i$ to their latest estimates, i.e.,\\
\begin{equation}
\begin{aligned}
\hat{\boldsymbol h}_i^{[m+1]}&=\arg \mathop {\max }\limits_{{\boldsymbol h_i}}\textrm{Re}\left\{\left(\boldsymbol y_i^{[m]}\right)^H\boldsymbol E\left(\hat{\varepsilon}_i^{[m]}\right)\boldsymbol F\hat{\boldsymbol X}_i^{[m]}\boldsymbol D\boldsymbol h_i\right\}\\
&{}=\boldsymbol K\boldsymbol D^H\left(\hat{\boldsymbol X}_i^{[m]}\right)^H\boldsymbol F^H\boldsymbol E^H \left(\hat\varepsilon_i^{[m]}\right)\boldsymbol y_i^{[m]},
\end{aligned}
\end{equation}
in which\\
\begin{equation}
\boldsymbol K=\left(\sigma_w^2\boldsymbol R_i^{-1}+\boldsymbol D^H\left(\hat{\boldsymbol X}_i^{[m]}\right)^H\hat{\boldsymbol X}_i^{[m]}\boldsymbol D\right)^{-1}
\end{equation}
and $\boldsymbol R_i$ is the covariance matrix of $\boldsymbol h_i$, $\boldsymbol R_i=E\left\{\boldsymbol h_i\boldsymbol h_i^H\right\}$. It is seen that (11) is equivalent to the MMSE estimation [12] obtained with the latest estimates of $\hat{\varepsilon}_i$ and $\hat{\boldsymbol X}_i$.

The frequency offset estimation at the $(m+1)$th iteration can be obtained by maximizing (9) with respect to $\varepsilon_i$ while keeping $\hat{\boldsymbol h}_i$ and $\hat{\boldsymbol X}_i$ fixed at their latest value, i.e.,\\
\begin{equation}
\hat{\varepsilon}_i^{[m+1]}=\arg \mathop {\max }\limits_{{\varepsilon_i}}\textrm{Re}\left\{\left(\hat{\boldsymbol y}_i^{[m]}\right)^H\boldsymbol E\left(\varepsilon_i\right)\boldsymbol F\hat{\boldsymbol X}_i^{[m]}\boldsymbol D\hat{\boldsymbol h}_i^{[m+1]}\right\}.
\end{equation}
To cope with the nonlinear problem in (13), we assume $\left|\varepsilon_i-\hat{\varepsilon}_i^{[m]}\right|$ is sufficient small such that we can replace $e^{j\frac{2\pi\varepsilon_i n}{N}}$ with its Taylor's series expansion around $\hat{\varepsilon}_i^{[m]}$ to the second order term, i.e.,\\
\begin{equation}
\begin{aligned}
e^{j\frac{2\pi\Delta \varepsilon _i^{[m]} n}{N}}{\approx} 1+j\frac{2\pi n}{N}\Delta \varepsilon _i^{[m]} +\frac{1}{2}\left(j\frac{2\pi n}{N}\right)^2\left(\Delta \varepsilon _i^{[m]}\right)^2,
\end{aligned}
\end{equation}
in which $\Delta \varepsilon _i^{[m]}=\varepsilon _i-\hat{\varepsilon} _i^{[m]}$. Substitute (14) into (13) and remove the terms that do not relate to $\varepsilon_i$, we have\\
\begin{equation}
\begin{aligned}
&\hat\varepsilon_i^{[m+1]}=\hat\varepsilon_i^{[m]}\\
&{}-\frac{N}{2\pi}\frac{\sum\limits_{n=0}\limits^{N-1}n\textrm{Im}\left\{\left(\hat{\boldsymbol y}_i^{[m]}\left(n\right)\right)^{\ast}\boldsymbol \Omega_i^{[m]}\left(n\right)\exp\left\{\frac{j2\pi\hat\varepsilon_i^{[m]}n}{N}\right\}\right\}}{\sum\limits_{n=0}\limits^{N-1}n^2\textrm{Re}\left\{\left(\boldsymbol y_i^{[m]}\left(n\right)\right)^{\ast}\boldsymbol \Omega_i^{[m]}\left(n\right)\exp\left\{\frac{j2\pi\hat\varepsilon_i^{[m]}n}{N}\right\}\right\}}.
\end{aligned}
\end{equation}
in which we let $\boldsymbol \Omega_i^{[m]}=\boldsymbol F\hat{\boldsymbol X}_i^{[m]}\boldsymbol D\hat{\boldsymbol h}_i^{[m+1]}$.
In order to update the value of $\boldsymbol X_i$, we replace equation (10) with\\
\begin{equation}
\boldsymbol y_i^{[m]}=\boldsymbol E\left(\varepsilon_i\right)\boldsymbol F\boldsymbol X_i\boldsymbol D\boldsymbol h_i+\boldsymbol I_{\bar i}^{[m]}+\boldsymbol w,
\end{equation}
in which $\boldsymbol I_{\bar i}^{[m]}=\boldsymbol E\left(\varepsilon_{\bar i}\right)\boldsymbol F\boldsymbol X_{\bar i}\boldsymbol D\boldsymbol h_{\bar i}-\boldsymbol E\left(\hat{\varepsilon}_{\bar i}^{[m]}\right)\boldsymbol F\hat{\boldsymbol X}_{\bar i}^{[m]}\boldsymbol D\hat{\boldsymbol h}_{\bar i}^{[m]}$ is the residual interference from $T_{\bar i}$ after the $m$th iteration. Note that $\boldsymbol I_{\bar i}^{[m]}\left(k\right)$ is a linear function of all symbols transmitted by $T_{\bar i}$. Therefore, it is rational to assume that $\boldsymbol I_{\bar i}^{[m]}$ is nearly Gaussian distributed with zero mean and covariance matrix $\sigma_I^2\boldsymbol I_N$ following the central limit theorem.
Then we obtain
\begin{equation}
\hat{\boldsymbol X}_i^{[m+1]}\left(k,k\right)=\arg\min\limits_{\boldsymbol X_i\left(k,k\right)}\left| \hat{\boldsymbol Y}_i\left(k\right) -\boldsymbol X_i\left(k,k\right)\hat{\boldsymbol H}_i\left(k\right) \right|^2,
\end{equation}
in which $\hat{\boldsymbol Y}_i=\boldsymbol A^T\boldsymbol F^H\boldsymbol E^H\left(\hat{\varepsilon}_i^{[m+1]}\right)\boldsymbol y_i^{[m]}$ and $\hat{\boldsymbol H}_i=\boldsymbol D\hat{\boldsymbol h}_i^{[m+1]}$.

Let $\hat{\boldsymbol \theta}=\left[\hat{\boldsymbol \varepsilon}^T, \hat{\boldsymbol h}^T, \hat{\boldsymbol X}^T\right]^T$ denote the final estimate of $\boldsymbol \theta=\left[{\boldsymbol \varepsilon}^T, {\boldsymbol h}^T, {\boldsymbol X}^T\right]^T$ after $M$ iterations. Then the ICI component in (3) can be reconstructed by\\
\begin{equation}
\hat{\boldsymbol I}_R=\left(\hat{\boldsymbol {\Pi}}_1-\hat{\boldsymbol {\Lambda}}_1\right)\hat{\boldsymbol S}_1+\left(\hat{\boldsymbol {\Pi}}_2-\hat{\boldsymbol {\Lambda}}_2\right)\hat{\boldsymbol S}_2,
\end{equation}
in which $\hat{\boldsymbol {\Pi}}_i=\boldsymbol F^H\boldsymbol E\left(\hat{\varepsilon}_i\right)\boldsymbol F$, $\hat{\boldsymbol {\Lambda}}_i=\textrm {diag}\left(\hat{\boldsymbol {\Pi}}_i\right)$ and $\hat{\boldsymbol S}_i=\hat{\boldsymbol X}_i\boldsymbol D\hat{\boldsymbol h}_i$.
\subsection{Channel-Decoding and Network-Coding Scheme}
In this subsection, we investigate the channel-decoding and network-coding scheme for frequency asynchronous PLNC. Notably, for synchronous PLNC, the channel-decoding and network-coding at the relay node consists of the following two steps [7].

\textit {Step-1}: In the first step, the relay maps received samples $\boldsymbol Y_R$ into the XORed massages $\boldsymbol b_1\oplus\boldsymbol b_2$ by function $T$. Specifically, the relay firstly computes the \emph{posteriori} probability $p\left(\boldsymbol u_1\left(k\right),\boldsymbol u_2\left(k\right)|\boldsymbol Y_R\left(k\right)\right)$ from the received samples. Then the log-likelihood ratios (LLRs) of network-coded information can be obtained by\\
\begin{equation}
\begin{aligned}
&\Phi\left(\boldsymbol c_1\left(l\right)\oplus\boldsymbol c_2\left(l\right)\right)=\\
&{}\log \left( {\frac{{\sum\limits_{\left(\boldsymbol u_1\left(k\right),\boldsymbol u_2\left(k\right)\right):\boldsymbol c_1\left(l\right)\oplus\boldsymbol c_2\left(l\right)=1} p\left(\boldsymbol u_1\left(k\right),\boldsymbol u_2\left(k\right)|\boldsymbol Y_R\left(k\right)\right) }}{{\sum\limits_{\left(\boldsymbol u_1\left(k\right),\boldsymbol u_2\left(k\right)\right):\boldsymbol c_1\left(l\right)\oplus\boldsymbol c_2\left(l\right)=0} p\left(\boldsymbol u_1\left(k\right),\boldsymbol u_2\left(k\right)|\boldsymbol Y_R\left(k\right)\right) }}} \right).
\end{aligned}
\end{equation}
Since with the same linear channel code at both source nodes, the XOR of two codewords $\boldsymbol c_1\oplus\boldsymbol c_2$ is also a valid codeword. Therefore, the relay can directly perform channel decoding over $\Phi\left(\boldsymbol c_1\left(l\right)\oplus\boldsymbol c_2\left(l\right)\right)$ to obtain $\boldsymbol b_1\oplus\boldsymbol b_2$.

\textit {Step-2}: In the second step, the relay re-channel encodes $\boldsymbol b_1\oplus\boldsymbol b_2$ and broadcasts the coded information in the BC phase.

\begin{figure}[t]
\centering
\subfigure[]{
\label{fig:subfig:a} %% label for first subfigu
\includegraphics[width=8.9cm]{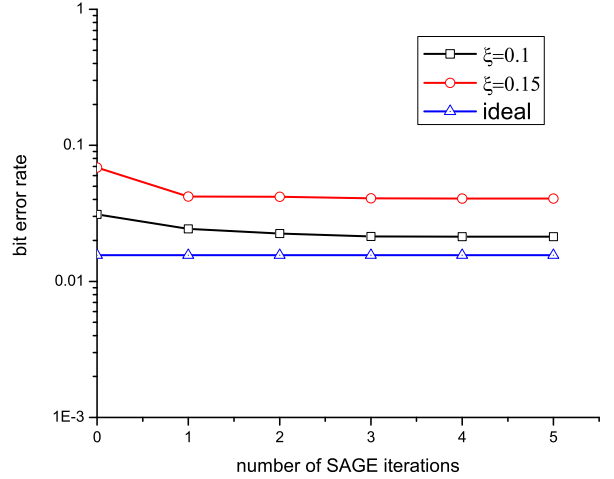}}
\hspace{1in}
\subfigure[]{
\label{fig:subfig:b} %% label for second subfigure
\includegraphics[width=8.9cm]{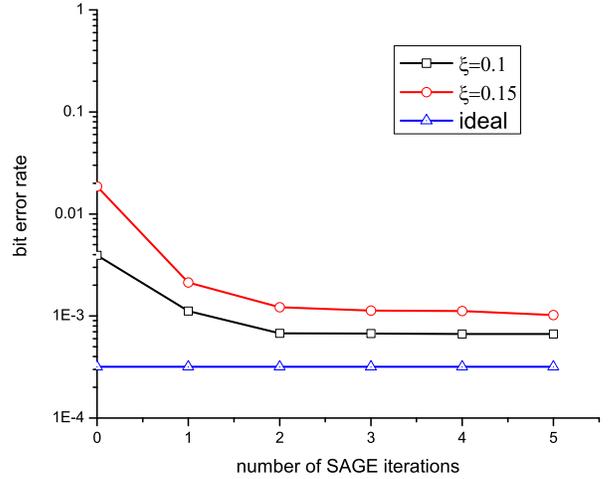}}
\caption{BER performance versus number of SAGE iterations for different normalized CFOs: (a) SNR=10dB; (b) SNR=15dB.}
\label{fig:subfig} %% label for entire figure
\end{figure}

For asynchronous case, we need to compute the \emph{posteriori} probability $p\left(\boldsymbol u_1\left(k\right),\boldsymbol u_2\left(k\right)|\boldsymbol Y_R, \boldsymbol{\varepsilon}, \boldsymbol h \right)$ in order to apply (18). However, the major difficulty occurs here is that it takes an exhaustive search over $2(N-1)$ dimensional space to compute this probability even for BPSK modulation with perfect knowledge of CIRs and CFOs. This is caused by correlations in the received samples. That is, due to the loss of orthogonality among subcarriers, each received symbol is affected by the interference from all other subcarriers as shown in (3). Consequently, each sample is correlated with all other samples.

To circumvent this obstacle, we present a three-step process to perform the channel-decoding and network-coding mapping at the node $R$.

\textit {Step-1}: The first step is referred to as the interference cancellation step. In this step, we intend to remove the ICI component in (3) using the reconstructed ICI presented in (18). By removing the ICI component from the frequency domain received samples, we obtain\\
\begin{equation}
{\boldsymbol{Y}_R'}=\boldsymbol A^T\left(\boldsymbol {\Lambda}_1\boldsymbol S_1+\boldsymbol {\Lambda}_2\boldsymbol S_2\right)+\boldsymbol A^T\left(\boldsymbol I_R-\hat{\boldsymbol I}_R\right)+\boldsymbol W,
\end{equation}
in which $\boldsymbol I_R=\left(\boldsymbol {\Pi}_1-\boldsymbol {\Lambda}_1\right)\boldsymbol S_1+\left(\boldsymbol {\Pi}_2-\boldsymbol {\Lambda}_2\right)\boldsymbol S_2$.

\textit {Step-2}: From (18), it is seen that $\hat{\boldsymbol I}_R$ is the estimate of ${\boldsymbol I}_R$. Here we suppose the elements of $\boldsymbol I_R-\hat{\boldsymbol I}_R$ or the estimation errors are sufficient small so that we can compute $p\left(\boldsymbol u_1\left(k\right),\boldsymbol u_2\left(k\right)|\boldsymbol Y_R, \boldsymbol{\varepsilon}, \boldsymbol h \right)$ approximately by\\
\begin{equation}
\begin{aligned}
&p\left(\boldsymbol u_1\left(k\right)=a,\boldsymbol u_2\left(k\right)=b|\boldsymbol Y_R, \boldsymbol{\varepsilon}, \boldsymbol h \right)\approx\\
&{}C\exp \left\{ { - \frac{1}{{\sigma _w^2}}{{\left| {\boldsymbol Y_R'\left( k \right) - a{\boldsymbol \Gamma _1}\left( k \right) - b{\boldsymbol \Gamma _2}\left( k \right)} \right|}^2}} \right\}
\end{aligned}
\end{equation}
in which $\boldsymbol \Gamma _i=\boldsymbol A^T \hat{\boldsymbol \Lambda}_i\boldsymbol D\hat{\boldsymbol h}_i$ and $C$ is a constant independent of $\boldsymbol u_1$ and $\boldsymbol u_2$. Then LLRs of the network-coded codewords could be computed by (18). After that, channel decoder is employed to map the LLRs into $\boldsymbol b_1\oplus\boldsymbol b_2$.

\textit {Step-3}: This step is identical with the \textit {Step-2} for synchronous PLNC.

\subsection{Complexity Analysis}
In this subsection, we study the computational complexity of the proposed scheme. Note that multiplications by matrices $\boldsymbol F$ and $\boldsymbol D$ are equivalent to DFT(IDFT) operations, which could be efficiently computed by FFT with $N{\log _2}N$ complex additions and ${N \mathord{\left/{\vphantom {N 2}} \right.\kern-\nulldelimiterspace} 2}{\log _2}N$ complex multiplications, respectively. Multiplications by matrices $\boldsymbol E\left(\hat{\varepsilon}_{i}\right)$ and $\hat{\boldsymbol X}_{i}$ require $N$ and $K$ complex multiplications, respectively. Therefore, it is shown that the total computational complexity for each SAGE iteration is $6N{\log _2}N+2N+K$ complex additions and $3N{\log _2}N+5N+K$ complex multiplications. Also, we can obtain that the computational load for (18)-(20) is $6N{\log _2}N+7K$ complex additions and $3N{\log _2}N+28K$ complex multiplications. According to the analysis above, it is seen that the overall complexity involved in the proposed two-step asynchronous PLNC scheme is approximately $\left(12M+6\right)N{\log _2}N+4MN+\left(2M+6\right)K$ complex additions and $\left(6M+3\right)N{\log _2}N+10MN+\left(2M+28\right)K$ complex multiplications, in which $M$ is the maximum number of SAGE iterations.

Notably, for scenarios that the CFOs are nearly constant, (15) can be computed only in the first block after the preamble to further reduce the computational complexity.
\section{Numerical Results}

\begin{figure}[t]
\centering
\includegraphics[width=8.9cm]{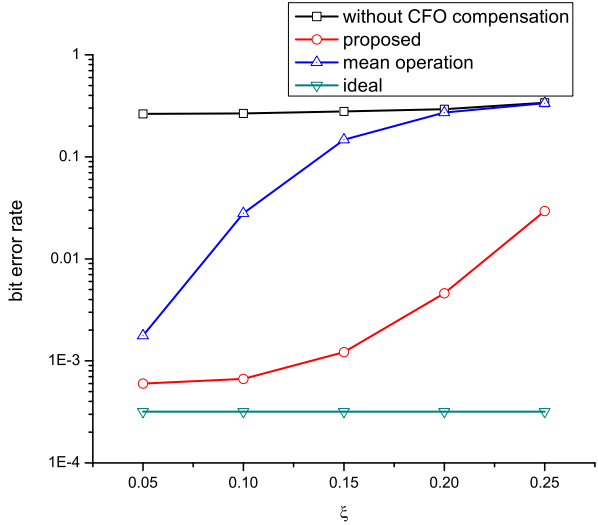}
\caption{BER performance versus normalized CFO, SNR is set to 15dB.} \label{fig:graph}
\end{figure}
\begin{figure}[t]
\centering
\includegraphics[width=8.9cm]{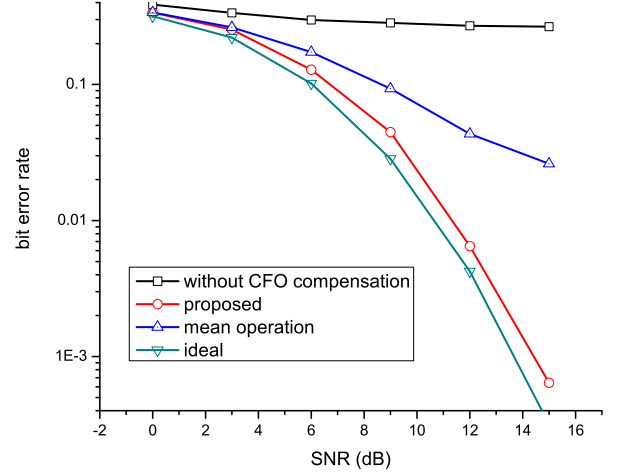}
\caption{BER performance versus SNR with constant CFOs in one uplink frame, where $\boldsymbol \varepsilon$ is set to $\boldsymbol \varepsilon=\left[0.1,-0.1\right]^T$.} \label{fig:graph}
\end{figure}

In this section, simulation results of the proposed scheme are presented. For the simulation setup, we consider a TWOR system with $N=128$ subcarriers. For simplicity, we allocate all the subcarriers to each terminal. BPSK modulation is assumed. Quasi-cyclic LDPC code [13] with codewords of length $1270$ and code rate $1/2$ is chosen and all nodes are assumed to use the same channel code. Channels between terminal nodes and relay node are modeled as six-tap frequency-selective fading and the power delay profile of CIR is presented as $E\left|h_i\left(l\right)\right|^2 \propto e^{-l/2}$ for $l=0,1,\cdots ,L_i-1$ and $\sum\limits_{l = 0}^{{L_i-1}} E{{{\left| {{h_i}\left( l \right)} \right|}^2} = 1}$.

It is assumed that the uplink frame of each terminal consists of 10 OFDM blocks. At the beginning of each frame, a preamble [10] is employed to estimate the CIRs which will be utilized to initial the iteration at the first OFDM block. The initial CFOs are set to $\hat{\boldsymbol\varepsilon}^{[0]}=\left[0,0\right]^T$. The CIRs are supposed to be constant in one uplink frame and the final estimates of $\boldsymbol\varepsilon$ and $\boldsymbol h$ at the last OFDM block are utilized to initial the next block.
\begin{figure}[t]
\centering
\includegraphics[width=8.9cm]{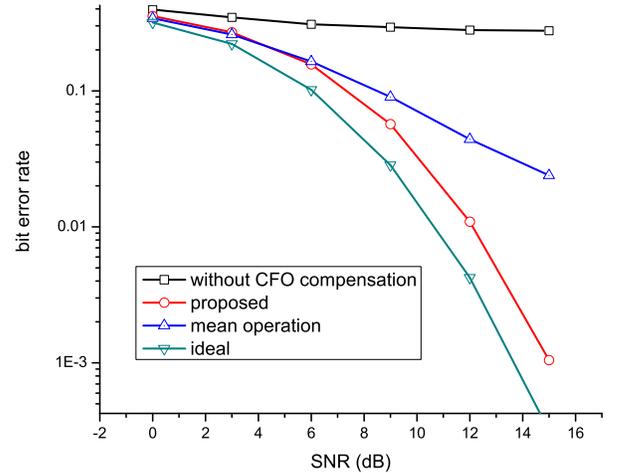}
\caption{BER performance versus SNR with time-varying CFOs in one uplink frame, where $\xi$ is set to 0.1.} \label{fig:graph}
\end{figure}

In Fig. 2, the BER performance versus number of SAGE iterations for different normalized CFOs is presented, where the SNRs are set to 10dB and 15dB. We set the CFOs as a function of $\xi$, that is, $\boldsymbol\varepsilon=\xi\cdot\left[-1,1\right]^T$ in which $\xi$ is modeled as a deterministic scalar belonging to interval $\left[0,0.5\right]$. The synchronous system with perfect knowledge of CIR is also considered to provide a benchmark. As shown in the figure, two iterations are sufficient for convergence of the proposed scheme. Hence, we fix the maximum number of SAGE iterations $M$ to $2$ in the rest of this section.

In Fig. 3, we present the BER performance of the proposed scheme as a function of normalized CFO. The normalized CFOs are also set as $\boldsymbol\varepsilon=\xi\cdot\left[-1,1\right]^T$ in which $\xi$ varies between -0.15 and 0.15. The compensation scheme proposed in [10] (This scheme is referred to as the mean operation.) and the synchronous system are also presented for comparison. As shown in the figure, the BER performance of both the proposed scheme and the mean operation degrades as the increase of normalized CFO. However, we can see that the proposed scheme remarkably outperforms the mean operation as well as the curve without CFO compensation.

In Fig. 4 and Fig. 5, the BER performance versus SNR for the proposed scheme is depicted. The CFOs are assumed to be constant during each uplink frame in Fig. 4. So the CFO estimation in (15) is operated only in the first block after the preamble. It is seen from the figure that the proposed scheme effectively mitigates the effect of frequency offsets in OFDM modulated PLNC. Particularly, the SNR loss is approximately 0.5dB at a BER of $10^{-3}$ for the case $\boldsymbol \varepsilon=\left[-0.1,0.1\right]^T$. In Fig. 5, we assume the CFO varies as a sinusoidal function of block index $n$ with an amplitude of $\% 5$ of the intercarrier spacing [14], i.e., $\varepsilon_i\left(n\right)=(-1)^{i}\xi + 0.05\sin\left(\frac{2}{5}\pi n\right)$, $n=1,2,\cdots,10$. Also, it can be observed that the proposed scheme remarkably outperforms the mean operation and the scheme without CFO compensation. The SNR loss is approximately 1.5dB at a BER of $10^{-3}$ for $\xi=0.1$.

In Fig. 6, we compare the BER performance of the proposed scheme with the mean operation for different relative CFOs. Here we define the relative CFO as $\left|\varepsilon_1-\varepsilon_2\right|$. Without loss of generality, we set $\varepsilon_1=0.05$ and let $\varepsilon_2$ vary between $-0.15$ and $0.15$. It is seen from the figure that BER performance of the mean operation deteriorates greatly as the relative CFO increases. However, our proposed scheme remarkably mitigates the performance degradation at the whole observation interval.

\section{Conclusions}
In this paper, we propose a two-step scheme to cope with the frequency asynchrony in TWOR. In the proposed scheme, SAGE algorithm is applied to reconstruct the ICI component from received signal at the relay. Then a channel-decoding and network-coding scheme is employed to map the received samples into the XOR of two terminals' information. It can be shown that the proposed scheme greatly mitigates the degradation due to CFOs with a relatively lower complexity and is robust to larger relative CFO comparing with the existing strategy.

\section*{Acknowledgment}
This work is supported by the Jiangsu Province Natural Science Foundation under Grant BK2011002, Major Special Project of China (2010ZX03003-003-01) and National Natural Science Foundation of China (No. 60972050).
\begin{figure}[t]
\centering
\includegraphics[width=8.9cm]{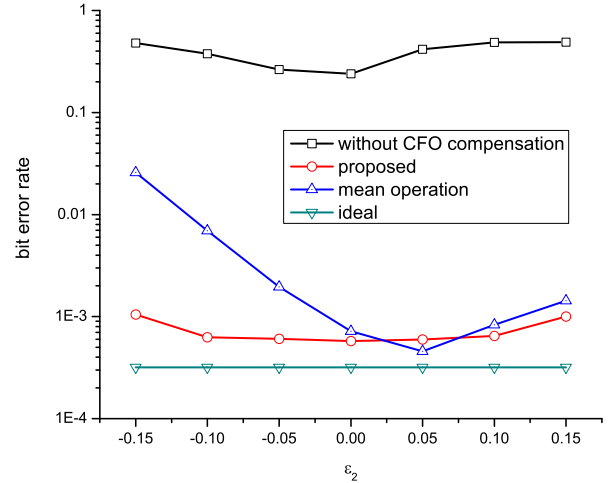}
\caption{BER performance versus normalized CFO, where $\varepsilon_1=0.05$ and SNR=15dB.} \label{fig:graph}
\end{figure}

\end{document}